\documentclass[letter]{aa} % for the letters 
\usepackage{graphicx}
\usepackage{txfonts}
\usepackage[colorlinks]{hyperref}
\hypersetup{
     colorlinks=true,
     linkcolor=blue,
     filecolor=blue,
     citecolor =blue,      
     urlcolor=blue,
     }
     
\usepackage{graphicx}	% Including figure files
\usepackage{amsmath}	% Advanced maths commands
\usepackage{amssymb}
\newcommand{\T}{$T_{\rm eff}$}
\newcommand{\g}{log($g$)}

\begin{document}

   \title{The chemical signature of the Galactic spiral arms \\
   revealed by Gaia DR3}
   
     \titlerunning{The chemical signature of the Galactic spiral arms  }

 % \subtitle{  }
 
%  \author{E. Poggio\inst{1,2} \and A. Recio-Blanco\inst{1}
%          \and P. A. Palicio\inst{1} \and P. Re Fiorentin\inst{2} \and P. de Laverny\inst{1}
%          \and R. Drimmel\inst{2} \and \\
%          G. Kordopatis\inst{1} \and
%          M. G. Lattanzi\inst{2} \and M. Schultheis\inst{1} \and A. Spagna\inst{2} \and E. Spitoni\inst{1} }

\author{E. Poggio\inst{1,2} \and A. Recio-Blanco\inst{1}
         \and P. A. Palicio\inst{1} \and P. Re Fiorentin\inst{2} \and P. de Laverny\inst{1}
         \and R. Drimmel\inst{2}\and \\
         G. Kordopatis\inst{1} \and
         M. G. Lattanzi\inst{2} \and M. Schultheis\inst{1} \and A. Spagna\inst{2} \and E. Spitoni\inst{1} }

  \institute{Université Côte d’Azur, Observatoire de la Côte d’Azur, CNRS, Laboratoire Lagrange, France\\
              \email{poggio.eloisa@gmail.com}
         \and
             Osservatorio Astrofisico di Torino, Istituto Nazionale di Astrofisica (INAF), I-10025 Pino Torinese, Italy\\
             }

  % \date{Received September 15, 1996; accepted March 16, 1997}

% \abstract{}{}{}{}{} 
% 5 {} token are mandatory
 
   \date{Received XXXX; accepted YYY}

  \abstract{ 
  %Taking advantage of the stellar parameters published in the \emph{Gaia} Data Release 3 (DR3), as well as astrometric measurements from \emph{Gaia} Early Data Release 3 (EDR3)
  
  Taking advantage of the recent \emph{Gaia} Data Release 3 (DR3), we map chemical inhomogeneities in the Milky Way's disc out to a distance of $\sim$ 4 kpc from the Sun, using different samples of bright giant stars. The samples are selected using effective temperatures and surface gravities from the GSP-Spec module, and are expected to trace stellar populations of different typical age. The cool (old) giants exhibit a relatively smooth radial metallicity gradient with an azimuthal dependence. Binning in Galactic azimuth $\phi$, the slope gradually varies from $d$[M/H]$/dR \sim -0.054$ dex kpc$^{-1}$ at $\phi \sim -20^{\circ}$ to $\sim -0.035$ dex kpc$^{-1}$ at $\phi \sim 20^{\circ}$. On the other hand, the relatively hotter (and younger) stars present remarkable inhomogeneities, apparent as three (possibly four) metal-rich elongated features in correspondence of the spiral arms' locations in the Galactic disc. When projected onto Galactic radius, those features manifest themselves as statistically significant bumps on top of the observed radial metallicity gradients with amplitudes up to $ \sim 0.05-0.1$ dex, making the assumption of a linear radial decrease not applicable to this sample. The strong correlation between the spiral structure of the Galaxy and the observed chemical pattern in the young sample indicates that the spiral arms might be at the origin for the detected chemical inhomogeneities. In this scenario, the spiral arms would leave in the younger stars a strong signature, which progressively disappears when cooler (and older) giants are considered. 
  
   %$\T \sim$ 5000 K and surface gravity \g \, < \, 1.5 dex.
  %($\T \sim$ 4000 K, \g < 1.5 dex)

  %The observed chemical inhomogeneities cause azimuthal variations of about $\lesssim 0.1$ dex in the hot RGB sample, and $\lesssim 0.05$ dex in the cool RGB sample.
  %However, the observed chemical patterns strongly depend on which sample is considered.
  
  %the extent and relative importance of the detected chemical inhomogeneities is larger for the of RGB stars containing hotter (i.e. larger effective temperature) stars. Specifically, the sample of hot RGB stars exhibit three, possibly four, elongated metal-rich features (stripes) in the Galactic plane, which are found to overlap with the position of the spiral arms in the Galactic disc. 
 %invalidating the usual adoption of a smooth decreasing profile.  suggesting a possible formation scenario for the observed chemical inhomogeneities.

 }

   \maketitle
%
%________________________________________________________________

\section{Introduction}

%
%                                            
%----------------------------------------------------------- 
   \begin{figure*}
   \centering
   \includegraphics[width=18cm]{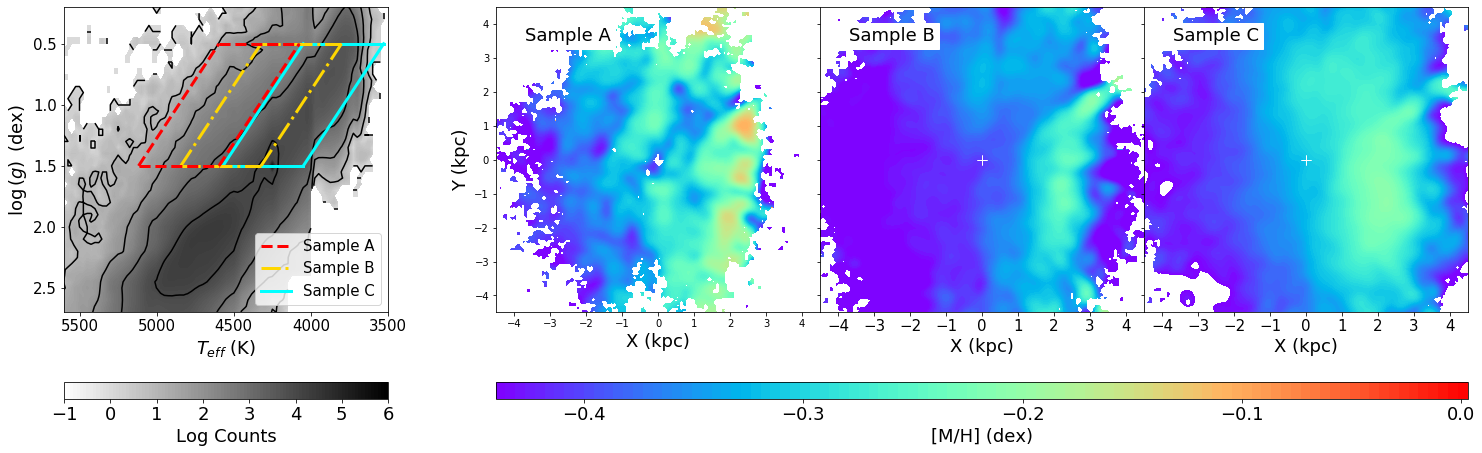}
      \caption{\emph{Chemical inhomogenities in the Galactic disc for three different samples.} \emph{Left panel}: Selection of Sample A, B and C in the Kiel diagram. \emph{Right panels}: Mean metallicity in heliocentric coordinates for Sample A, B, and C (from left to right{\bf, respectively}) in the Galactic plane. The position of the Sun is shown by the white cross in (X,Y)=(0 kpc, 0 kpc); the Galactic center is to the right; Galactic rotation is going clockwise.
              }
         \label{fig:selection}
   \end{figure*}
%______________________________________________________________

It has been known for several decades that the disc of the Milky Way contains large-scale non-axisymmetric features, including the spiral arms and a central bar \citep[see e.g.][and references threin]{Georgelin:1976, Okuda:1977, Shen:2020, DR3-DPACP-75}, which can play an important role in the disc’s evolution, and are expected to leave their signature in the observed properties of the stars. 

%A great revolution is changing the understanding of our home Galaxy, the Milky Way. After centuries of effort, the picture of the Milky Way is now snapping into focus. The main reason for this revolutionary change is 

%For instance, it has been known for several decades that the disc contains large-scale non-axisymmetric features, including the spiral arms and a central bar \citep[see e.g.][ XXX add other references]{DR3-DPACP-75}, which can play an important role in the disc’s evolution, and are expected to leave their signature in the observed properties of the stars. 

%In disk galaxies, spiral arms are expected to manifest their influence in several ways, 

Several models have tried to explain the behaviour of spiral arms in disc galaxies, although their dynamical nature and physical origin still remain unknown. Spiral arms have been proposed to induce large-scale shock waves that trigger the gravitational collapse of gas clouds, enhancing star formation \citep{Roberts:1969}. They can also induce radial migration of stars \citep{Lynden-Bell:1972, Sellwood:2002, Schonrich:2009a, Schoenrich:2009b, Roskar:2012}, and trap or scatter stars close to orbital resonances \citep{Contopoulos:1986}. As a consequence, they can generate variations in the mean chemical composition of stars in the arm and inter-arm regions \citep{Minchev:2012, Kordopatis:2015, Grand:2016, Khoperskov:2018, Spitoni:2019, Khoperskov:2021}. %Mapping chemical inhomogeneities in the Galactic disk can, therefore, potentially constrain the nature 

Mapping the spatial distribution of metals within galaxies therefore represents a key aspect to study the processes of chemical enrichment and mixing in the interstellar medium. An extensive review of the chemical enrichment of galaxies across the cosmic epochs can be found in \citet{Maiolino:2019} (see also references therein). In external galaxies, several works have found evidence of azimuthal variations in the mean chemical composition of stars \citep[e.g.][]{SanchezMenguiano:2016, Ho:2017, Vogt:2017}. Recently, using data from the PHANGS–MUSE survey \citep{Emsellem:2022}, \citet{Williams:2022} found that most of the star-forming galaxies considered in their sample showed significant 2D metallicity variations.

In the Milky Way, chemical inhomogenities were explored by several studies. \citet{Balser:2011} analyzed the oxygen distribution of HII regions in the Galactic disc, divided their sample into three Galactic azimuth bins, and found that the corresponding  radial gradients were significantly different. 
%alayzing Cepheids in the Galactic disc...
Chemical inhomogeneities in iron and oxygen were also found in Cepheids \citep{Pedicelli:2009,Genovali:2014,Kovtyukh:2022}.
Large metallicity variations were also found in the interstellar medium \citep{DeCia:2021}. By combining data from HII regions, Cepheids, B stars and red supergiants, \citet{Davies:2009} presented evidence for large azimuthal variations ($\sim$ 0.4 dex) in oxygen, magnesium and silicon on a scale of few kpc in the inner Galactic disc. On the other hand, using APOGEE red-clump stars, \citet{Bovy:2014} constrained azimuthal variations in the median metallicity to be $\leq$ 0.02 dex in the region covered by their sample. Using the large spectroscopic survey RAVE and the Geneva Copenhagen Survey, \citet{Antoja:2017} analyzed stars in a cylinder of 0.5 kpc radius centered on the Sun, and found asymmetric metallicity patterns in velocity space.

%understanding the nature of spiral structure in disk galaxies is one of the main, and still unanswered questions in galactic astronomy

%tackle outstanding questions about physical mechanisms regulating the evolution of the Milky Way’s disc.

%The great wealth of new information, eventually complemented by ground-based surveys (e.g. WEAVE, GALAH, 4MOST, MOONS…), will be crucial in the coming years to derive a multidimensional view of the Galaxy, and ultimately understand the processes that shaped its formation and evolution.
%Gaia data, in synergy with ground-based surveys (e.g. WEAVE, GALAH, 4MOST, MOONS…), are and will be crucial in the coming years to derive a multidimensional view of the Galaxy, and ultimately understand the processes that shaped its formation and evolution.

%By studying the galaxy HCG91c, \citet{Vogt:2017} detected azimuthal variations the oxygen-abundance gradient tracing the spiral arms. Based on observations from external spiral galaxies, it has been suggested that the enrichment of the interstellar medium can proceeded preferentially along spiral structures, and less efficiently across them \citep{Vogt:2017}.   In our Milky Way, Gaia DR3 now allows us to map and quantify chemical inhomogeneities in the Galactic disc, and estabilish which mechanism(s) might be at work.

%The \emph{Gaia} satellite from the European Space Agency (ESA) is currently measuring photometry, positions, and motions for more than one billion stars with exquisite accuracy and unparalleled detail. 

%homogeneous
Recently, \emph{Gaia} Data Release 3 \citep[hereafter DR3, ][]{DR3-DPACP-185} published the largest stellar catalog with chemical abundances, atmospheric parameters and radial velocities ever created. Radial Velocity Spectrometer (RVS) data were parameterised by the General Stellar Parameteriser - spectroscopy (\emph{GSP-Spec}) module \citep{GSPspec_RecioBlanco2022}, delivering chemo-physical parameters for 5.6 million stars over the entire sky. \citet{DR3-DPACP-104} observed azimuthal variations in red giant branch stars, but didn't present quantitative results or in-depth discussions on their origin.
%In \citet{DR3-DPACP-104}, azimuthal variations in red giant branch stars were observed, but neither quantitative results nor in-depth discussions were presented.

%, both in terms of size and sky-coverage. 
%This allows us to link 
%this new information is linked to positions and motions from previous \emph{Gaia} releases \citep[see ][]{Brown:2021} over the entire sky. 
Thanks to the large number of chemical and astrometric measurements, as well as \emph{Gaia}'s all-sky sampling, it is now possible to map chemical inhomogeneities in the Milky Way's disc as never before, with the goal of detecting the possible signature left by the Galactic spiral arms in the stellar metallicity.

%Previous \emph{Gaia} releases delivered positions, and motions for more than one billion stars with exquisite accuracy \citep[see ][]{Brown:2021, Brown:2018}, allowing us to link chemical abundances of stellar atmospheres, stellar positions and motions over the entire sky for the first time. 
%Thanks to the great wealth of new available information, it is now possible to study with unprecedented detail the processes regulating the formation and evolution of the Galactic disc. 

%Thanks to the great wealth of new available information, 
The Letter is structured as follows: in Section \ref{Sec:data} we describe the datasets; in Section \ref{Sec:results} we present our results; in Section \ref{Sec:conc} we discuss our findings and future perspectives.

\section{Data selection} \label{Sec:data}

In this study, we use the main stellar atmospheric parameters (effective temperature \T, surface gravity \g, global metallicity [M/H]) 
derived from \emph{Gaia} RVS spectra by the \emph{GSP-Spec} module, to select tracers of the disc population and map large-scale chemical inhomogeneities in our Galaxy. 
The quality constraints on spectroscopic and astrometric measurements (as well as a Toomre kinematical selection) are detailed in Appendix \ref{appendix_cuts}. 

%To map chemical inhomogeneities in the Galactic disc, we use stellar parameters from the GSP-Spec module (cit GSP spec paper) published as part of Gaia Data Release 3. 
%In this study, we use stellar atmospheric parameters and metallicities derived from \emph{Gaia} RVS spectra by the General Stellar Parameteriser - spectroscopy (GSP-Spec) module (XXXXX add citation). We select only stars with 5-parameters astrometric solution, \verb+ruwe+ <1.4, and satisfying quality flags on astrometric and spectroscopic measurements, as well as a Toomre kinematical selection to remove possible halo contaminants (details are given in Appendix \ref{appendix_cuts}). Finally, GSP-Spec estimates of effective temperature \T\ and surface gravity \g\ are used to select and compare different samples.
To serve our purpose, we focus on giant stars, 
%as targets to map large-scale inhomogeneities in the Galactic disc, 
as they are intrinsically bright, and therefore allow us to sample a relatively large volume of the Galactic disc. Also, taking advantage of the large number of high-quality chemical measurements for these stars\footnote{As an indication, for stars with \g $<$ 2.5 and $6000 <$ \T $< 3500$ K, the metallicity uncertainty is $<$ 0.1 dex for 1.4 million stars, $<$ 0.05 dex for $\sim$680 000 stars, and $<$ 0.01 dex for $\sim$43 000 stars.}, a robust statistical analysis can be performed, thanks to high signal-to-noise spectra.

%Moreover, we note that a large number of high-quality chemical measurements and 
%stellar parameters was published for RGB stars in Gaia DR3,

%as this would allow us to obtain statistically robust results

%need intrinsically bright objects, so that large distances from the Sun can be reached. . The above mentioned requirements suggest that the red giant branch (RGB) stars can be considered a suitable stellar population. The portion of the Kiel diagram occupied by RGB stars is shown in Figure \ref{fig:selection}, left panel.For the purposes of this work, red giant branch (RGB) stars can be considered suitable stellar tracers, because: (i) they are intrinsically bright, and therefore can be mapped out to large distances; (ii) compared to the brightest stars on the upper main sequence, the quality of the chemical measurements is higher, and they are more numerous, allowing better statistics. 

Figure~\ref{fig:selection} (left panel) shows the portion of the Kiel diagrams populated by the giant stars in our selected sample. Here we %focus on 
select stars brighter than the red clump (which is located at approximately \g $\simeq $ 2.3 and \T $\simeq $ 4750 K), to avoid selection function effects due to the superposition of different stellar populations \citep[see Section 4 of ][]{DR3-DPACP-104}. Indeed, such effects can cause artefacts when investigating radial gradients, which might be erroneously interpreted as signatures of the spiral arms. Selected areas in Fig.~\ref{fig:selection} (left panel) define three different subsamples (details in Appendix \ref{appendix_cuts}), which cover different portions of the red giant branch. Additionally, we apply a vertical cut $|Z|= | d \sin{b} | < 0.75$ kpc, where $b$ is the galactic latitude and $d$ is the heliocentric distance according to the geometric model by \citet{BailerJones:2021}. The adopted samples are here labelled as A, B, and C, and contain respectively 19\,340, 151\,139, and 346\,570 stars. 

The portion of the Kiel diagram covered by our three samples is expected to be populated by stars of different ages. Both theoretical considerations (based on isochrones) and empirical evidence (based on spatial distribution and stellar kinematics) indicate that Sample A is expected to be typically younger than the other two samples (more details in Appendix \ref{appendix_sample_characterization}). For instance, the comparison between the selected areas on the Kiel diagram and the PARSEC isochrones \citep{Bressan:2012,Chen:2014,Chen:2015,Tang:2014,Pastorelli:2019} for metallicities between $[M/H]=$ 0 dex and $[M/H]=$-0.5 dex suggests that Sample A should mostly contain stars younger than 100-300 Myr (taking into account typical uncertainties on \T~ and \g), while Sample C should be dominated by much older stars, i.e. $\gtrsim$ 1 Gyr (see Fig.~\ref{fig:isochrones}). On the other hand, Sample B is expected to be intermediate between the two, containing a mixture of young and old stellar populations. As will be shown in the following, the different content of the three samples will determine remarkable differences in their observed metallicity maps. 

%Presumably, this is the main reason why Sample A, B and C exhibit remarkable differences in the observed chemical maps, as will be shown in the following Section.

%Sample A is expected to contain stars typically bluer than the ones in Sample C (dynamically hotter), while Sample B is expected to be somewhat intermediate. The content of the three samples is further explored in Appendix \ref{appendix_sample_characterization}.

%Similar to the selection performed in \cite{DR3-DPACP-104}, here we adopt an inclined slope of $0.00192$ K$^{-1}$, so that it approximately follows the natural inclination of the RGB in the Kiel diagram. (sentence on the width).

%Sample C is very similar to the RGB sample used in \cite{DR3-DPACP-104}.

\section{Results} \label{Sec:results}
 
Figure \ref{fig:selection} (right panels) shows the maps of the mean metallicity $\langle \, $[M/H]$ \, \rangle $ in the Galactic plane for Samples A, B and C, using a smoothing Gaussian kernel with a bandwidth of 175 pc (see Appendix \ref{appendix_bivariate}). %(Details on the bivariate smoothing used for the maps can be found in Appendix \ref{appendix_bivariate}). 
As discussed in the following, the three maps exhibit both similarities and differences. On a large scale, stars are typically more metal-rich toward the inner parts of the Galaxy for all three samples. This is in agreement with previous observations of radial metallicity gradients in the Galactic disc \citep[e.g.][]{Genovali:2014,DR3-DPACP-104}, and also expected from the inside-out formation scenario \citep{bird2013,matteucci2021}. However, the maps also present smaller-scale metallicity inhomogenehities. The relative importance, extent, and shape of the observed inhomogeneities vary for the three different samples. Sample A exhibits three, possibly four, metal-rich elongated features, that diagonally cross the portion of the XY-plane covered by our dataset. One feature extends approximately from (X,Y)=($-$1.5~kpc, 3~kpc) to ($-$3~kpc, $-$1~kpc). A second feature stretches from (X,Y)=(1~kpc, 3.5~kpc) to (X,Y)=($-$1.5~kpc, $-$1~kpc), and then continues almost vertically out to (X,Y)=($-$1.5~kpc, $-$3~kpc). A third feature (which possibly contains multiple segments) extends from (X,Y)=(2.5~kpc, 1~kpc) to (X,Y)=(1~kpc, $-$4~kpc). This pattern becomes progressively less evident as we move from Sample A to Sample B and C. In Sample C, the elongated features disappear almost completely, and simply result in an observed asymmetry about the Sun of stars at $X \simeq$ 0 kpc, typically more metal rich at $Y > 0$ kpc than at $Y < 0$ kpc. While such asymmetry was already noted by \citet{DR3-DPACP-104} in their RGB sample, the scenario presented in this contribution indicates that it might represent a remnant of the spiral arms chemical signature.

% which gradually becomes more (less) evident when a sample having similar to Sample A (Sample C) is considered.

%; the scenario presented by this work points to that asymmetry as a remnant of the signature  in the scenario presented in this work, given the comparison with Sample B and Sample A,  in the scenario presented by this contribution, such asymmetry 

%directly points to the spiral arms.

%In this work, Sample B 

%______________________________________________________________
\begin{figure*}
\centering
\includegraphics[width=18cm,trim={5cm 1.5cm 3.5cm 2.5cm},clip]{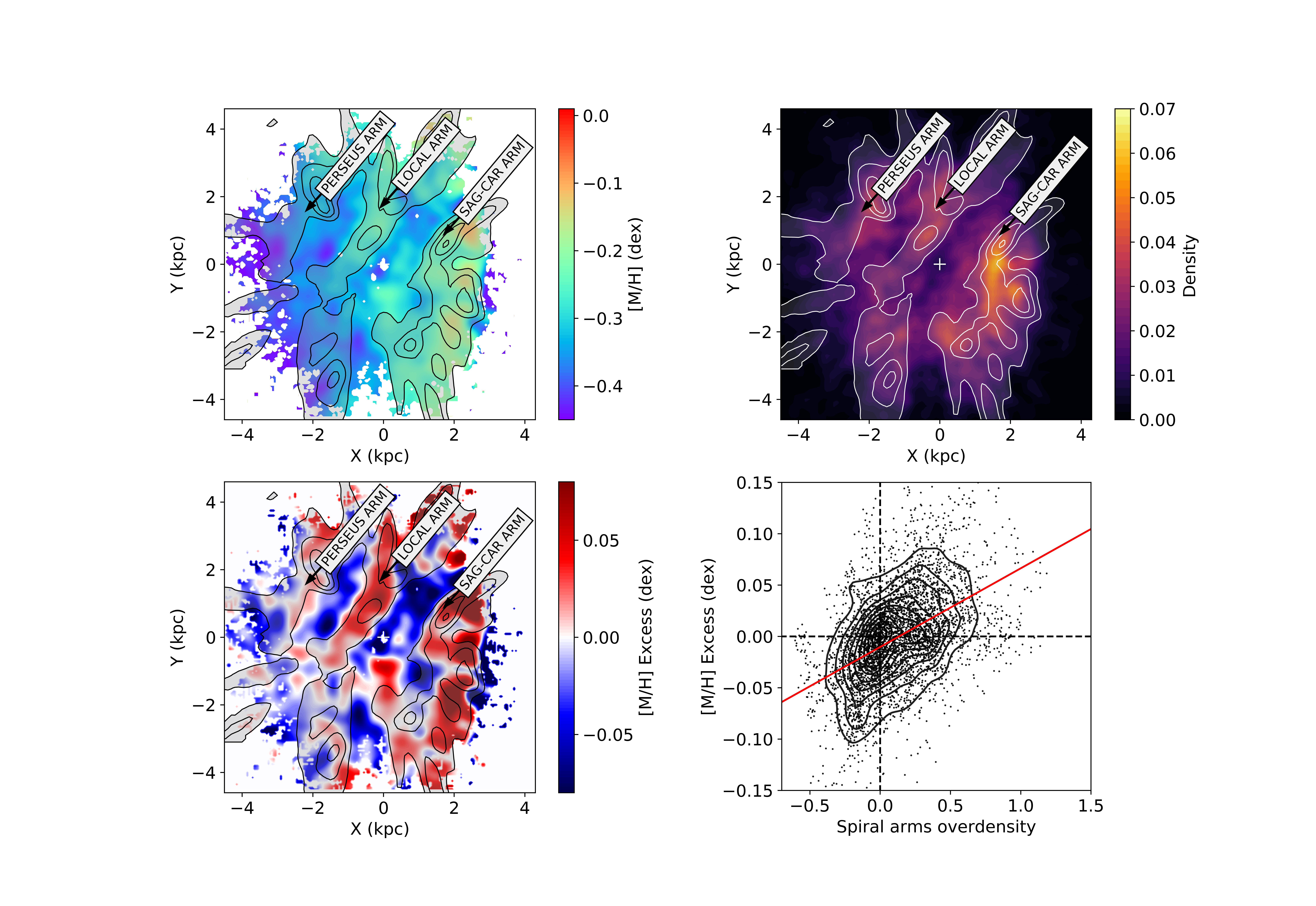}
\caption{ \emph{Correlation between the observed metal-rich features in Sample A and the segments of the nearest spiral arms in the Galaxy.} \emph{Upper left} panel: Map of the mean metallicity in the Galactic plane for Sample A shown in Fig.~\ref{fig:selection}, but now compared to the position of the spiral arms from Upper Main Sequence stars (UMS) in \citet{Poggio:2021}, shown as grey shaded areas. The labels show the position of the Perseus arm, the Local (Orion) arm and the Sagittarius-Carina arm based on the UMS map. \emph{Upper right panel}: bivariate kernel density estimation for Sample A using an Epanechnikov kernel with a smoothing length
of 250 pc. The contours are the same as the upper left panel. \emph{Lower left panel}: Same as Upper panels, but now showing the metallicity excess. Regions with positive (negative) metallicity excess are more metal-rich (-poor) than the average (see text for a detailed explanation). \emph{ Lower right panel}: Using the map shown in the lower left panel, the overdensity of the spiral arms vs. the metallicity excess is shown for each pixel (black dots). When the spiral arm overdensity is $>0$, the region can be considered belonging to a spiral arm, whereas regions with $<0$ can be considered interarm regions. Black contours show the distribution of the pixels in the spiral arms overdensity vs. metallicity excess space. The red line shows a linear fit to the points.
\label{fig:corr}
}
\end{figure*}
%______________________________________________________________

Figure \ref{fig:corr} (upper left panel) shows again the mean metallicity for Sample A, but now compared to the position of the spiral arms in the Galaxy as mapped by Upper Main Sequence UMS stars in \citet{Poggio:2021}, shown as grey shaded areas and black contours, indicating (from left to right) the segments of: the Perseus arm, the Local (Orion) arm, and the Sagittarius-Carina (Sag-Car) arm (the latter contour possibly containing the Scutum arm).

In the upper right panel of Fig.~\ref{fig:corr}, we can see the spatial distribution of sources in Sample A, obtained with an Epanechnikov kernel density estimator with a bandwidth of 250 pc. White contours show again the position of the nearest spiral arms from UMS stars. As we can see, Sample A traces well the position of the spiral arms, confirming that it contains typically young stars.

To better analyse and map chemical inhomogeneities in the Galactic disc, we define a new variable, called Metallicity Excess, defined as $ \langle \, \rm{[M/H]} \, \rangle_{\rm{loc}} - \langle \, \rm{[M/H]} \, \rangle_{\rm{large}} $, where $\langle \, [\rm{M/H}] \, \rangle_{\rm{loc}}$ and $ \langle \, [\rm{M/H}] \, \rangle_{\rm{large}} $ represent the mean metallicity smoothed on a local or large scale, respectively. The smoothing is performed as described in Appendix \ref{appendix_bivariate}. The lower left panel of Fig.~\ref{fig:corr} shows the obtained Metallicity Excess for Sample A if we adopt for the local smoothing the same value chosen for the mean metallicity in the upper left panel of Fig.~\ref{fig:corr} (i.e. $ \langle \, [\rm{M/H}] \, \rangle_{\rm{loc}} = \langle \, [\rm{M/H}] \, \rangle$), and for the large-scale metallicity a scale-length of 5 times the value adopted for the local smoothing. The red (blue) regions in the Metallicity Excess map should be interpreted as places where the stars are more metal-rich (-poor) than the average.

We therefore study the correlation between the observed metal-rich regions and the geometry of the spiral arms by dissecting the XY map (e.g. lower left panel of Fig.~\ref{fig:corr}) in pixels of 0.3 kpc width. For each pixel, we consider the spiral arms overdensity %(calculated as described in P21)
\citep[calculated as described in][]{Poggio:2021}
and the Metallicity Excess. The bottom right panel of Fig.~\ref{fig:corr} shows the correlation between these two quantities (every black dot corresponds to a pixel in the XY map). The red line shows a linear fit to the black dots. As we can see, regions located in correspondence of the spiral arms (i.e. with overdensity larger than 0) tend to be more metal-rich than the average (i.e. with a Metallicity Excess larger than 0). To quantify the correlation between these two variables, we calculate Kendall's correlation coefficient $\tau_B=0.38$, which can be considered as an indication of strong positive correlation\footnote{A correlation is usually considered very weak for values less than 0.10,  weak between 0.10 and 0.19, moderate between 0.20 and 0.29, strong for 0.30 or above \citep{botsch2011chapter}.}. Therefore, the elongated metal-rich features observed in Sample A appear to be statistically correlated with the position of the spiral arms in the Galaxy.

It should be noted, however, that there are some exceptions. For instance, the enhancement of metallicity at approximately (X,Y)=( 0 kpc, $-$1 kpc) does not seem to correspond to an overdensity in the spatial distribution. The comparison between our map and the data from \citet{Hottier:2021} reveals that it might correspond to the Vela Molecular Ridge. This region can be seen as a dust enhancement in the maps of \citet{Lallement:2019,Vergely:2022}, and possibly prevents us from seeing an overdensity in the spatial distribution of stars (for both Sample A and UMS stars, see the upper right panel in Fig.~\ref{fig:corr}).

%the lack of observed stars in both the UMS and Sample A spatial distribution ().  } %some regions of the Galactic plane are more metal rich than the average, but they do not coincide with the spiral arms contours. {\bf The enhancement of metallicity at (X,Y)=(0 kpc, $-$1 kpc that does not seem to correspond to an over density of OB stars } (e.g. (X,Y)=(0 kpc, $-$1 kpc, {\bf which might correspond to the Vela Molecular Ridge }) in Fig.~\ref{fig:corr}, upper and bottom left panel). 
Furthermore, it is interesting to note that not all the regions in the spiral arms exhibit the same chemical behaviour: the metallicity excess seems to be typically high in the Sag-Car arm and in the upper part (Y>0) of the Local arm, but seems to be milder in the lower part of the Local arm (Y<0) and the Perseus arm (Y $\lesssim$ 1 kpc).

\begin{figure}
\centering
\includegraphics[width=6.5cm]{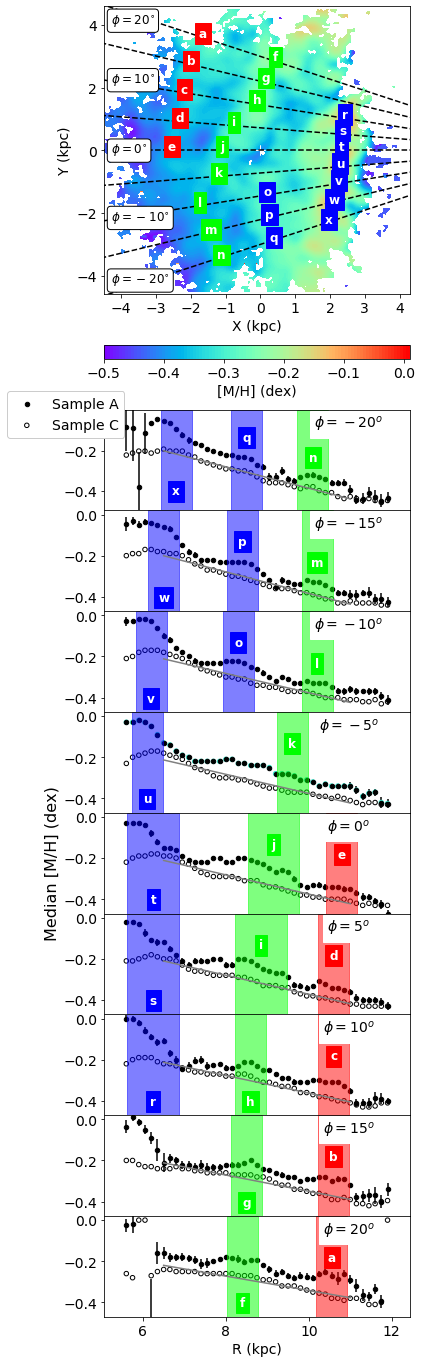}
\caption{ \emph{Impact of the observed chemical inhomogeneities on the Galactic gradients.} \emph{Upper panel}: Map of the mean metallicity in the Galactic plane for Sample A, with overlaid azimuthal slices of 5$^{\circ}$, from $\phi = -20^{\circ}$ to $\phi = 20^{\circ}$.  \emph{Lower panel}: median metallicity as a function of Galactocentric radius $R$ for different slices in Galactic azimut $\phi$ for Sample A ({\bf filled} dots) and Sample C (open dots). Regions that might be associated with the Perseus (red, letters `a'--`e'), Local (green, `f'--`n'), and Sagittarius/Scutum (blue, `o'--`x') arms are shown as colored regions both in the upper and lower panels.
\label{fig:projection_radial_gradient}
}
\end{figure}

Figure~\ref{fig:projection_radial_gradient} shows the impact of the above-described metal-rich features on the observed metallicity gradient in the Galactic disc. To explore the metallicity variations in the disc, we dissect the Galactic plane into slices of 10$^{\circ}$ width in Galactic azimuth $\phi$. For each slice, we then show the median metallicity for Sample A and Sample C. Sample B presents an intermediate behaviour between Sample A and C (here not shown to preserve the clarity of the Figure). As we can see, Sample A presents some peaks of higher metallicity, superimposed on a global decrease as a function of $R$, here calculated assuming a distance to the Galactic center $R_{\odot}=8.249$~kpc \citep[][]{Gravity20}. Based on bootstrap uncertainties, the peaks deviate more than 3-sigma from a linear decrease as a function of Galactic radius, indicating that the radial metallicity gradient is dominated by chemical undulations, that reach maximum deviations of $\sim$ 0.05-0.1 dex. In this context, the observed peaks can be interpreted as the projection on the radial direction of the metal-rich features observed in the XY maps. For Sample C, on the other hand, a relatively smooth decrease of the median metallicity as a function of $R$ is apparent. The slopes of the radial metallicity gradients for different azimuthal bins can be found in Table~\ref{tab:slope}. As we can see, the radial metallicity gradient gradually varies depending on azimuth, and becomes gradually steeper for $\phi < 0^{\circ}$.

%Add this: -> More than 3-sigma deviation from a linear gradient?
Additional tests to verify the robustness of our results can be found in Appendix \ref{appendix_additional_tests}.

%______________________________________________________________
%______________________________________________________________

%                                             Simple A&A Table
%_____________________________________________________________
%
\begin{table}
\caption{Radial gradient for Sample C in different azimuthal bins, as shown in Fig.~\ref{fig:projection_radial_gradient}, assuming a linear model. Only stars with $R>6.5$~kpc are here considered, to avoid selection effects in the inner regions due to extinction, as well as $R < 11.5$~kpc, to avoid low statistics regions. The corresponding table for Sample A and B is not presented here due to the spiral arms-related metallicity variations, which are invalidating the assumption of a linear model. \label{tab:slope} }             % title of Table
\label{table:1}      % is used to refer this table in the text
\centering                          % used for centering table
\begin{tabular}{r c c}        % centered columns (4 columns)
%\hline\hline                 % inserts double horizontal lines
\hline\hline  
%& [M/H] &\\    % table heading 

$\phi$ (deg) & Slope (dex kpc$^{-1}$) & Intercept (dex) \\    % table heading 
\hline\hline                       % inserts single horizontal line
$-$20 $\pm$ 5 & $-$0.054 $\pm$ 0.001 & 0.15 $\pm$ 0.01 \\ 
$-$15 $\pm$ 5 & $-$0.052 $\pm$ 0.002 & 0.14 $\pm$ 0.01 \\   
$-$10 $\pm$ 5 & $-$0.047 $\pm$ 0.002 & 0.09 $\pm$ 0.02 \\   
$-$5 $\pm$ 5 & $-$0.046 $\pm$ 0.002 & 0.09 $\pm$ 0.01 \\   
0 $\pm$ 5 & $-$0.047 $\pm$ 0.001 & 0.09 $\pm$ 0.01 \\   
5 $\pm$ 5 & $-$0.046 $\pm$ 0.001 & 0.09 $\pm$ 0.01 \\   
10 $\pm$ 5 & $-$0.043 $\pm$ 0.001 & 0.06 $\pm$ 0.01 \\   
15 $\pm$ 5 & $-$0.039 $\pm$ 0.001 & 0.04 $\pm$ 0.01 \\   
20 $\pm$ 5 & $-$0.035 $\pm$ 0.002 & 0.01 $\pm$ 0.01 \\   
\hline                                   %inserts single line
\end{tabular}
\end{table}
%
%_____________________________________________________________

%and are more prominent for Sample A, as expected. However, the amplitude and location of the observed peaks depends on how the rings are taken with respect to the spiral arms.

\section{Discussion and Conclusion} \label{Sec:conc}

% ----> Add this to the discussion: <-----
%Sample A contains an excess of young stars
%which is presumably the cause for the differences in the chemical inhomogeneities mapped in ...

Taking advantage of the new \emph{Gaia} DR3 data, we have mapped the mean metallicity in the Galactic disc out to 3-4 kpc of the Sun for three different samples of bright giant stars. We found evidence of chemical inhomogeneities, which manifest themselves as statistically significant undulations on top of the expected radial metallicity gradients. The appearance and magnitude of the detected inhomogeneities are different for the three considered samples, being more structured and pronounced for the sample containing hotter (and a larger relative fraction of young) stars. In this sample, three (possibly four) extended metal rich features are detected, which appear to be statistically correlated with the position of the spiral arms (independently derived from previous works). Such connection points to the spiral arms as a plausible origin for the observed metal-rich features. 

It is interesting to compare our maps of the mean metallicity in the Milky Way's disc {to the results based on} external spiral galaxies. For the galaxy HCG91c, \citet{Vogt:2017} found that the chemical enrichment of the interstellar medium proceeds preferentially along spiral structures, and less efficiently across them, in agreement with our results. \citet{Williams:2022} considered a sample of 19 galaxies, for which detected significant chemical variations in most cases, but found no clear signs of enrichment along the spiral arms. On the other hand, \citet{Sanchez:2020} detected the presence of more metal-rich HII regions in the spiral arms with respect to the corresponding interarm regions for a large subsample of galaxies, in agreement with the maps presented here for the Milky Way.

 While the observed inhomogeneities are more pronounced and structured for the youngest stars, azimuthal variations in the old stellar populations also deserve some discussion. The oldest stars considered here (Sample C) present a metallicity gradient that changes with Galactic azimuth, with a slope that varies from $\sim -$0.054 dex kpc$^{-1}$ to $\sim -$0.035 dex kpc$^{-1}$. To our knowledge, this is the first time that such azimuthal dependence is shown for old stellar populations, both in the Milky Way and in external galaxies. The slopes obtained here for Sample C using different azimuthal bins are in good agreement (considering the uncertainties) with those from the giant stars at all azimuths in the Galactic plane in \citet{DR3-DPACP-104} (see their Table 1), which have a slope of $ -$0.055 $\pm$ 0.007 dex kpc$^{-1}$ above the plane, and $ -$0.057 $\pm$ 0.007 dex kpc$^{-1}$ below the plane. Previous works reported slopes for thin disc (i.e. low $\alpha -$sequence) field stars ranging from $\sim -$0.053 dex kpc$^{-1}$ to $\sim -$0.068 dex kpc$^{-1}$ \citep{Bergemann:2014,RecioBlanco:2014,Hayden:2015,Anders:2017}.

Recently, \citet{Hawkins:2022} mapped azimuthal variations in the Galactic disc, using both \emph{Gaia} DR3 data and a sample OBAF-type stars from the LAMOST survey published by \citet{Xiang:2022}. Using the LAMOST sample, the author found azimuthal structure on top of the radial gradient, which doesn't necessarily follow the spiral arms. On the other hand, using \emph{Gaia} DR3, \citet{Hawkins:2022} found good agreement with the results presented in this \emph{Letter}, finding that giant stars near the spiral arms typically contain more metals than those in the interarm regions. The author suggests that the discrepancy between the two samples can be due to the fact that either LAMOST data is too local to see the overall structure, or that azimuthal structure depends strongly on age, selection function and spectral type of the tracer population.

Several works available in the literature can offer a theoretical framework for the obtained observational results. Using high-resolution $N$~-body simulations, \cite{Khoperskov:2018} showed that kinematically hot and cold stellar populations in the Galactic disc react in a different way to a spiral arm perturbation, naturally leading to azimuthal variations in the mean metallicity of stars in the simulated disc \citep[based on the assumption that younger stars typically tend to have higher metallicity and smaller random motions than the older stellar components, see for example][]{Holmberg:2007}. Figure~4 in \citet{Khoperskov:2018} shows the azimuthal metallicity variations at different times of the simulation for three different initial metallicity profiles. They found that the spiral arms locii typically tend to be more metal-rich than the interarm-regions. %The coicidence between 
%However, the metal-rich feature do not always con
%possibly with smaller-scale variations, depending on the adopted model. 
Their predictions are in good agreement with the observations presented in this work. 

%According to their proposed scenario, the chemical signature of the spiral arms found in this contribution can be seen as the joint action of both kinematically hot and kinematically cold stellar populations in the disc. In this context, Sample A would show the signature better than the other two because it contains a larger relative fraction of (young) kinematically cold stars. %, with a more balanced proportion between the two different stellar populations.

%It is also worth noting that metallicity azimuthal variations were predicted by cosmological simulations. \citet{Grand:2016} found that metal-rich star particles (originated from internal regions of the disc) are transported outward along the trailing edge of the spiral, while the metal-poor particles (originated from the outer disc regions) are transported radially inwards along the leading edge of the spiral. Hence, as a result of the radial migration caused by the spiral arms, the residual metallicity pattern is systematically more metal-rich (-poor) along the trailing (leading) edge of the spiral arm at many Galactocentric radii. 

In the future, a detailed modelling will be crucial to fully explain the observed chemical inhomogeneities in the Galactic disc. Chemo-dynamical evolution models should be adopted to interpret the impact of the spiral arms on chemical variations in the Galactic disc, and to understand how different mechanisms can influence the observed present-day chemical abundance patterns \citep[e.g.][]{Roskar:2012,Grand:2016,Minchev:2018,Spitoni:2019,Spitoni:2022a,Spitoni:2022b,Carr:2022}. Future works comparing observations from \emph{Gaia} DR3 and ground-based surveys to sophisticated theoretical models will be able to show further insights on the evolutionary processes that shaped the present-day appearance of the Milky Way's disc.

%   \begin{enumerate}
%       \item The conditions for the stability of static, radiative
%          layers in gas spheres, as described by Baker's (\cite{baker})
%          standard one-zone model, can be expressed as stability
%          equations of state. These stability equations of state depend
%          only on the local thermodynamic state of the layer.
%       \item If the constitutive relations -- equations of state and
%          Rosseland mean opacities -- are specified, the stability
%          equations of state can be evaluated without specifying
%          properties of the layer.
%       \item For solar composition gas the $\kappa$-mechanism is
%          working in the regions of the ice and dust features
%          in the opacities, the $\mathrm{H}_2$ dissociation and the
%          combined H, first He ionization zone, as
%          indicated by vibrational instability. These regions
%          of instability are much larger in extent and degree of
%          instability than the second He ionization zone
%          that drives the Cephe{\"\i}d pulsations.
%   \end{enumerate}

\begin{acknowledgements}
The authors thank the referee for comments and suggestions
that improved the overall quality of the paper. This work has made use of data from the European Space Agency (ESA)
mission Gaia (https://www.cosmos.esa.int/gaia), processed by the Gaia Data Processing and Analysis Consortium (DPAC, https://www.cosmos.
esa.int/web/gaia/dpac/consortium). Funding for the DPAC has been provided by national institutions, in particular the institutions participating in the Gaia Multilateral Agreement. EP acknowledges support by the Centre National d'études Spatiales (CNES). ARB and ES received funding from the European Union’s Horizon 2020 research and innovation program
under SPACE-H2020 grant agreement number 101004214 (EXPLORE project). 

\end{acknowledgements}

%-------------------------------------------------------------------

% BIBLIOGRAPHY
% - use BibTeX with the regular commands:
   \bibliographystyle{aa} % style aa.bst
   \bibliography{mybib} % your references Yourfile.bib
%
% - join the .bib files when you upload your source files
%-------------------------------------------------------------------

\begin{appendix}

\section{Data quality cuts and selection\label{appendix_cuts}}

We select Gaia DR3 sources with atmospheric parameters (available through the 
%{\it astrophysical$_{-}$parameters} table)
\verb+gaiadr3.astrophysical_parameters+ table), radial velocities, and 5-parameters astrometric solution having quality index \verb+ruwe+ $< 1.4$ \citep{Lindegren2018}. In addition, \emph{Gaia} duplicated sources are excluded from the sample. As for the \emph{GSP-Spec} parameters, we reject objects based on their metallicity uncertainty and \emph{GSP-Spec} quality flags \citep{GSPspec_RecioBlanco2022} reporting on the degree of biases from line broadening and radial velocity errors affecting [M/H], flux noise, extrapolation on stellar atmospheric parameters, as well as KM-type stars.

The definition of the working sample is as follows:

%\bigskip
\begin{verbatim}
 [M/H]_unc<0.5 & vbroadM<2 & vradM<2 &
 fluxNoise<4 & extrapol<3 & KMtypestars<2 
 astrometric_params_solved = 31 & ruwe < 1.4 
 & duplicated_source == false 
\end{verbatim}

%where the cuts on the astrometric parameters are performed to select only stars with 5-parameter astrometric solutions and 
% where $[M/H]_{unc}$ is the uncertainty on GSP-Spec metallicity, $vbroadM$ and $vradM$ are ....,
% The condition {\it astrometric$_{-}$params$_{-}$solved = 31} implies that only 5-parameter astrometric solutions are retained, and the cut Renormalised Unit Weight Error (RUWE)
%which concern both GSP-Spec parameters and astrometric measurements. 
From the Gaia archive the following query provides the data set employed: %\begin{verbatim} gaiadr3.astrophysical_parameters \end{verbatim}
%GSP-Spec sample (containing 5 591 594 stars), we select only those 
%avaiable on the \emph{Gaia} archive, we select the objects that satisfy the following cuts:
%\bigskip

%OLD QUERY
% SELECT g*., ap.teff_gspspec, 
% ap.logg_gspspec, ap.mh_gspspec 
% FROM gaiadr3.gaia_source AS g 
% INNER JOIN gaiadr3.astrophysical_parameters AS ap 
% ON g.source_id = ap.source_id 
% WHERE ( 
% ((mh_gspspec_upper-mh_gspspec_lower)<0.25) 
% AND ((flags_gspspec LIKE "__0%") 
% OR (flags_gspspec LIKE "__1%")) 
% AND ((flags_gspspec LIKE "_____0%") 
% OR (flags_gspspec LIKE "_____1%")) 
% AND ((flags_gspspec LIKE "______0%") 
% OR (flags_gspspec LIKE "______1%") 
% OR (flags_gspspec LIKE "______2%") 
% OR (flags_gspspec LIKE "______3%")) 
% AND ((flags_gspspec LIKE "_______0%") 
% OR (flags_gspspec LIKE "_______1%") 
% OR (flags_gspspec LIKE "_______2%")) 
% AND ((flags_gspspec LIKE "____________0%") 
% OR (flags_gspspec LIKE "____________1%")) 
% ) AND g.astrometric_params_solved== 31 
% AND g.ruwe < 1.4

%\begin{lstlisting}[caption={\texttt{ADQL} query for the sample}.},captionpos=b]

%FIXED ISSUES ON THE QUERY
\noindent \begin{verbatim}

SELECT g.*, ap.teff_gspspec, 
ap.logg_gspspec, ap.mh_gspspec 
FROM gaiadr3.gaia_source AS g 
INNER JOIN gaiadr3.astrophysical_parameters AS ap 
ON g.source_id = ap.source_id 
WHERE ( 
((mh_gspspec_upper-mh_gspspec_lower)<0.25) 
AND ((flags_gspspec LIKE `__0%') 
OR (flags_gspspec LIKE `__1%')) 
AND ((flags_gspspec LIKE `_____0%') 
OR (flags_gspspec LIKE `_____1%')) 
AND ((flags_gspspec LIKE `______0%') 
OR (flags_gspspec LIKE `______1%') 
OR (flags_gspspec LIKE `______2%') 
OR (flags_gspspec LIKE `______3%')) 
AND ((flags_gspspec LIKE `_______0%') 
OR (flags_gspspec LIKE `_______1%') 
OR (flags_gspspec LIKE `_______2%')) 
AND ((flags_gspspec LIKE `____________0%') 
OR (flags_gspspec LIKE `____________1%')) 
) AND g.astrometric_params_solved = 31 
AND g.ruwe < 1.4
\end{verbatim}
%\end{lstlisting}

Additionally, objects without $d_{Geo}$ distances were discarded. \\

%Disk selection:
For the obtained sample, we calculate the Galactocentric coordinates $R, \phi, Z$ and the corresponding velocities $V_R, V_{\phi}, V_Z$ following the same conventions and parameters adopted in \cite{DR3-DPACP-104}. We apply a cut |Z|<0.75 kpc, as we are mainly interested in selecting disc stars. To this end, we also apply a kinematical cut based on the Toomre diagram \citep[see for example][]{Bensby:2003,Bensby:2014, ReFiorentin:2019, ReFiorentin:2021, DR3-DPACP-104, Goldensample:2022} in order to remove possible halo contaminants:
%select a sample of \emph{bona fide} thin disk stars:

\begin{equation}
    \sqrt{V_R^2 + V_Z^2 + (V_\phi - V_0)^2} < 210 \,\,\, \mbox{km~s}^{-1},
\end{equation} 
where $V_0=238.5$~km~s$^{-1}$ is the velocity of the local standard of rest at the Sun's position, based on  \citet{Gravity20} and \citet{SchBinDeh2010}.

%X, Y, Z positions, and cylindrical radius R) and
%the Galactocentric cylindrical velocities (VR, Vϕ and VZ in
%a right-handed frame with Vϕ positive for most of the disc
%stars), we adopted the Sun’s Galactocentric position (R, Z)⊙ =
%(8.249, 0.0208) kpc (Gravity Collaboration et al. 2021; Bennett
%& Bovy 2019) and Galactocentric cylindrical velocities$(VR, Vϕ, VZ)⊙ = (−9.5 

%RGB selection:
Finally, we select different populations of RGB stars in the \g-\T\ plane of the Kiel diagram. This is accomplished 
%Finally, as described in Section \ref{Sec:data} of this paper, we apply a selection on the Kiel diagram, 
by selecting stars in the three boxes shown in the left panel of Fig.~\ref{fig:selection}, that is:

\bigskip

\noindent \begin{verbatim}
logg < 1.5 & logg > 0.5 &
(logg > (coeff*teff + interc_left)) &
(logg < (coeff*teff + interc_right)) 
\end{verbatim}

\bigskip

where \verb+coeff+= 0.00192 $\rm{ dex \, K}^{-1}$ is the adopted slope, chosen to follow the natural inclination of the RGB branch \citep[as done also for the selection of the massive sample in][]{DR3-DPACP-104}, whereas the intercept of the two inclined lines, delimiting the selected regions on the left and right side, are: \verb+interc_left+ = -8.3 + $\Delta$ dex and \verb+interc_right+ = -7.3 + $\Delta$ dex, where $\Delta$ = 0, 0.5 and 1 for Sample A, B and C, respectively.

%______________________________________________________________
%______________________________________________________________

\section{Sample characterization \label{appendix_sample_characterization}}
%______________________________________________________________
\begin{figure*}
\centering
\includegraphics[width=13cm]{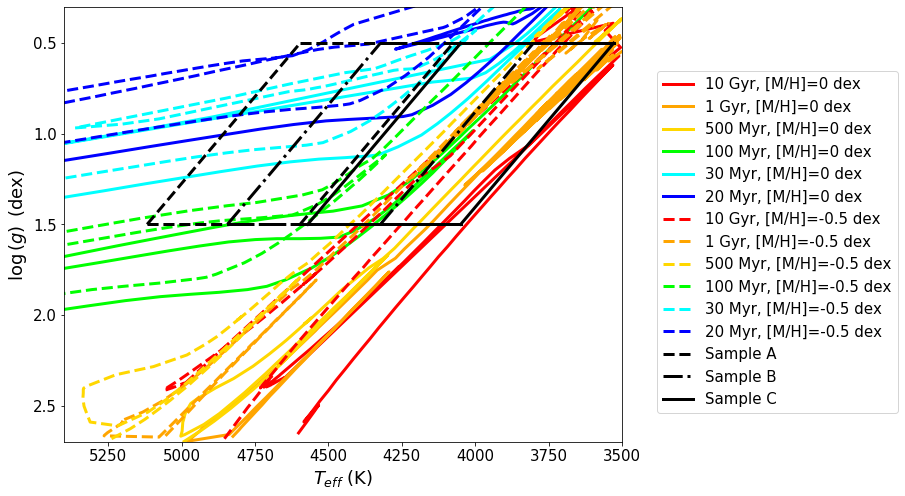}
\caption{ PARSEC isochrones for different metallicites and ages, compared to the regions of the Kiel diagram selected here for Sample A, B and C. \label{fig:isochrones}
}
\end{figure*}
%______________________________________________________________

%______________________________________________________________
\begin{figure}
\centering
\includegraphics[width=8cm]{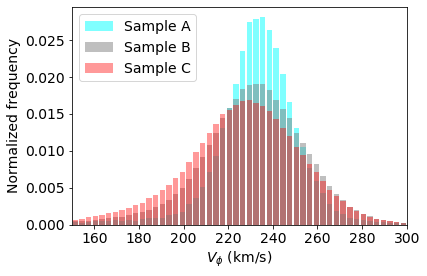}
\caption{ Distribution of azimuthal velocities $V_{\phi}$ for Sample A, B, and C. The histogram is normalized so that the area under the histogram integrates to 1 for all three samples. \label{fig:characterization}
}
\end{figure}
%______________________________________________________________

Here we performed some tests to further explore the content of Sample A, B and C. First of all, we compared our selected areas of the Kiel diagram with the prediction from the PARSEC isochrones for different metallicities \citep{Bressan:2012,Chen:2014,Chen:2015,Tang:2014,Pastorelli:2019}. Figure \ref{fig:isochrones} shows the isochrones for 10 Gyr, 1 Gyr, 500 Myr, 100 Myr, 30 and 20 Myr, for both solar metallicity [M/H]=0 dex and [M/H]=-0.5 dex. It should be noted that effective temperatures from \emph{GSP-Spec} have been found to be in agreement with those from the isochrones \citep[see Section 10.2 of ][]{GSPspec_RecioBlanco2022}, and no important bias as been found with the literature. As we can see in Fig~\ref{fig:isochrones}, in the region covered by Sample A, only isochrones for 100 Myr or younger are present. Of course, in real data, we do expect some contamination from older stars, given that uncertainties will tend to blur the distribution. Nevertheless, the fact that Sample B and C contain stars typically older than those in Sample A is supported by the location of the isochrones in the Kiel diagram.

Further indications can be also found empirically.
Although ages can be also inferred for stars \citep[e.g.][]{Kordopatis:2022}, here we simply use stellar kinematics as a proxy for the typical age of the sample. For Sample A, B and C, respectively, the velocity dispersions are: $\sigma_{V_Z}=(19.5, 21.8, 23.6)$~km~s$^{-1}$, $\sigma_{V_R}=(28.6, 35.1, 38.9)$~km~s$^{-1}$,
$\sigma_{V_{\phi}}=(26.6, 26.9, 29.0)$~km~s$^{-1}$. Figure \ref{fig:characterization} shows the distribution of the azimuthal velocities $V_{\phi}$. As we can see, the distribution of stars in Sample A presents a prominent peak at high azimuthal velocity ($\sim$ 230-240~km~s$^{-1}$). On the contrary, Sample B and Sample C present a broader distribution, with a large fraction of stars at low $V_{\phi}$, as would be expected as a consequence of the asymmetric drift for a kinematically hot stellar population. 

%As an additional test, we cross-matched Sample A, B and C with the catalog recently presented by \cite{Kordopatis:2022} (in prep), selecting only stars with relative age error < 0.5. As expected, we found that the stars younger than 200 Myr are preferentially located in proximity of the spiral arms \citep[using the map from ][]{Poggio:2021}. Additionally, we found that Sample A contains a relative fraction of young stars larger than the other two samples (for stars with age less than 300 Myr, it is 15 times larger than for Sample B, and 60 times larger than for Sample C). 

%The above-described tests, togeethboth support the scenario that Sample A contains an excess of young (and kinematically cold) stars compared to the other two samples. A discussion on this possibly explains the observed chemical maps is presented in Sec. \ref{Sec:conc}.

%The resulting distribution in age and initial mass is shown in Figure \ref{fig:characterization} (upper and lower panel, respectively). Based on the age estimates from \cite{Kordopatis:2022}, the relative fraction of young stars is larger for Sample A than the other two samples. On the other hand, Sample B and Sample C contain a larger fraction of low-mass (and older) stars, which might explain why the chemical signature of the spiral arms is less evident in those two samples. 

%______________________________________________________________
%______________________________________________________________

\section{Bivariate smoothing \label{appendix_bivariate}}
For a given position (X,Y) in the Galactic plane, we calculate the local smoothed metallicity:
\begin{equation}
    \langle \, {\rm[M/H]} \, \rangle_{\rm loc} \, (X,Y)= \frac{\sum^N_i {\rm[M/H]}_i \, K\left( \frac{X-X_i}{h_X}\right) \,  K\left(\frac{Y-Y_i}{h_Y}\right)}{\sum^N_i K\left( \frac{X-X_i}{h_X}\right) \,  K\left(\frac{Y-Y_i}{h_Y}\right)} \quad,
    \label{equation:smoothing}
\end{equation} 
where $N$ is the total number of stars in the sample, $X_i$ and $Y_i$ are the X- and Y-coordinates of the $i$-th star of the sample, $K$ is a Gaussian kernel 
\begin{equation}
      K\biggl( \frac{X-X_i}{h_X} \biggr) = \frac{1}{\sqrt{2 \pi} \, h_X} \exp{ \Biggl( -\frac{1}{2} \Biggl( \frac{(X-X_i)}{h_X}  \Biggr)^2 \Biggr)}
\end{equation} 
with a scale-length $h_X$ (and similarly for the Y-coordinate). In this work, local smoothed metallicity (see Fig. 1, right panels) is shown using a scale-length of 0.175 kpc. As discussed in the main text, the metallicity excess is calculated by taking the difference of the local metallicity (i.e. Equation \ref{equation:smoothing}, assuming a scale-length of 0.175 kpc) and the metallicity on a large scale (i.e. again using Equation \ref{equation:smoothing}, but now adopting a scale-length 5 times larger than the one used for the local metallicity).

\section{Azimuthal variations \label{appendix_azimuthal_variations}}

\begin{figure}
\centering
\includegraphics[width=6.5cm]{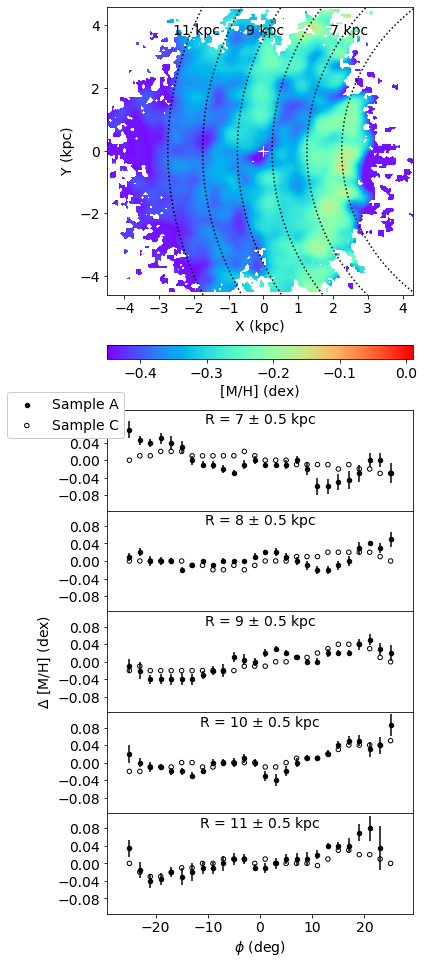}
\caption{ \emph{Impact of the observed chemical inhomogeneities on azimuthal variations.} \emph{Upper panel}: Map of the mean metallicity in the Galactic plane for Sample A, with overlaid rings of constant Galactocentric radius at $R= 6,7,8,9,10$ and $11$~kpc. \emph{Lower panels}: median metallicity for sample A and C (filled/open dots, respectively) as a function of Galactic azimuth $\phi$ for different rings in $R$ (as specified in each panel), after the median metallicity of the stars for each ring is subtracted. 
\label{fig:azimuthal_variations}
}
\end{figure}

Figure \ref{fig:azimuthal_variations} shows a projection of the observed metal-rich spiral features when the Galactic plane is cut into Galactocentric rings. For each ring, we show the azimuthal variations for both Sample A and Sample C, calculated as the median metallicity as a function of azimuth, after subtracting the median metallicity of the sample in each ring. We note that the azimuthal variations depend on the adopted sample, being more prominent for Sample A; typically, they do not exceed 0.08 dex. The amplitude and location of the observed peaks depend on how the rings are selected with respect to the spiral arms' geometry.

%______________________________________________________________
%______________________________________________________________

\section{Additional tests \label{appendix_additional_tests}}
{\bf Here} we present some additional tests that we perform to verify the robustness of our results.

First, we checked the quality of the astrometric measurements in our selected stars. For Sample A, 99.7 \% of the stars have a parallax signal-to-noise greater than 5. For those stars, as a further check, we constructed a map using distances inferred by the inverse of the parallax, obtaining consistent results.

As an additional test, we selected only the stars satisfying the criteria on \emph{Gaia} DR3 astrophysical parameters listed for the high-quality sample in \citet{DR3-DPACP-104}. In this case, the chemical signature of the spiral arms is still present, but the entire map is shifted toward more metal-rich values, as expected (given that more metal-rich stars are more likely to be included in the high-quality sample).
%as their spectral have more lines

To test the possible contribution from Asymptotic Giant Branch (AGB) stars, we cross-matched our three samples with the long-period variables catalog available in \emph{Gaia} DR3 \verb+gaiadr3.vari_long_period_variable+. We found that, for all three samples, the percentage of AGB stars from that catalog is less than 0.2 \%.

Our maps are also quite robust against the selection of the regions in the Kiel diagram. Indeed, a change in the shape of the selection leave the maps almost unchanged - although, of course, a significant shift in \T\ (or \g) can make an impact, as discussed in the main text.  

Finally, we test other spiral arms maps. The spiral arms contours used in the main text are from \citet{Poggio:2021}, which relies on distances calculated as described in \citet{Poggio:2018}. As a test, we also used the spiral arms overdensity map from \citet{DR3-DPACP-75}, where the $d_{Geo}$ distances are adopted from \citet{BailerJones:2021}, as well as the one based the sample from \citet{Zari:2021} (which used astro-kinematic distances), always obtaining consistent results.

\end{appendix}

\end{document}